\documentclass[aps,pra,floatfix,reprint,showpacs,longbibliography]{revtex4-1}

\usepackage{graphicx} 
\usepackage{dcolumn} 
\usepackage{bm} 
\usepackage{amssymb}
\usepackage{rotating} 
\usepackage{epstopdf}
\usepackage{epsfig}
\usepackage[caption=false]{subfig} 
\usepackage{amsmath}
\usepackage{color}

\begin{document}

\title{Non-equilibrium steady state conductivity in cyclo[18]carbon and its boron nitride analogue }
\author{Alexandra E Raeber and David A Mazziotti}
\email{damazz@uchicago.edu}
\affiliation{Department of Chemistry and The James Franck Institute, The University of Chicago, Chicago, IL 60637 USA}
\date{Submitted August 6, 2020}
\begin{abstract}
A ring-shaped carbon allotrope was recently synthesized for the first time, reinvigorating theoretical interest in this class of molecules. The dual $\pi$ structure of these molecules allows for the possibility of novel electronic properties.  In this work we use reduced density matrix theory to study the electronic structure  and conductivity of cyclo[18]carbon and its boron nitride analogue, B\textsubscript{9}N\textsubscript{9}. The variational 2RDM method replicates the experimental polyynic geometry of cyclo[18]carbon. We use a current-constrained 1-electron reduced density matrix (1-RDM) theory with Hartree-Fock molecular orbitals and energies to compute the molecular conductance in two cases: (1) conductance in the plane of the molecule and (2) conductance around the molecular ring as potentially driven by a magnetic field through the molecule's center.  In-plane conductance is greater than conductance around the ring, but cyclo[18]carbon is slightly more conductive than B\textsubscript{9}N\textsubscript{9} for both in-the-plane and in-the-ring conduction.   The computed conductance per molecular orbital provides insight into how the orbitals---their energies and densities---drive the conduction.
\end{abstract}

\maketitle

\section{Introduction}

 The structure and properties of ring-shaped carbon allotropes, known as cyclo[n]carbons, have been of theoretical interest for many years, first appearing in the literature more than fifty years ago~\cite{hoffmann1966extended}. As a result of their sp hybridization, cyclo[n]carbons have two orthogonal $\pi$ systems, allowing for the possibility of double aromatic stabilization~\cite{fowler2009double, baryshnikov2019cyclo}.  Their high reactivity, however, means that their synthesis was not possible until very recently. Last year, the first of these carbon allotropes, cyclo[18]carbon, was synthesized and characterized using high-resolution atomic force microscopy~\cite{kaiser2019sp}.

Prior to its synthesis, two possible structures had been hypothesized for cyclo[18]carbon, a cumulenic form, without bond length alternation, and a polyynic form, with alternating single and triple bonds. The cumulenic structure, which is predicted by H\"{u}ckel theory, is generally found to be the lowest in energy by density functional theory (DFT)~\cite{hutter1994structures, saito1999second, neiss2014nature} and M{\o}ller-Plesset second-order perturbation theory~\cite{parasuk199118} calculations. The polyynic structure, on the other hand is supported by Hartree-Fock~\cite{diederich1989all, plattner1995c18}, Monte-Carlo~\cite{torelli2000electron}, and coupled cluster calculations~\cite{arulmozhiraja2008ccsd}. Recent experimental work confirmed the polyynic geometry to be correct~\cite{kaiser2019sp, scriven2020synthesis, hussain2020vibrational}.

Boron nitride (BN) forms hexagonal lattices which are isoelectric to those formed by carbon. Its two-dimensional hexagonal lattice is analogous to graphene and has high thermal and chemical stability~\cite{strout2001structure}. Theoretical work on boron nitride analogues to cyclo[n]carbons has shown that they may have similar novel properties with higher stability~\cite{martin1996structure, giuffreda2000structural, pichierri2020boron}.

Here we study the electronic structure and molecular conductivity of the cyclo[18]carbon and B\textsubscript{9}N\textsubscript{9} molecules using reduced density matrix (RDM) theory~\cite{mazziotti2004first,mazz2004, gidofalvi2008active, mazz2011, mazz2012sec,sajjan2018current, raeber2019current}.  Computations with the variational two-electron reduced density matrix (2-RDM) method~\cite{mazziotti2004first,mazz2004, gidofalvi2008active, mazz2011, mazz2012sec}, which can treat fractionally filled orbitals and strong electron correlation if present~\cite{schlimgen2016entangled, montgomery2018strong, boyn2020entangled}, reveal that neither molecule is strongly correlated and support the experimental polyynic structure of cyclo[18]carbon.  Using a current-constrained 1-electron reduced density matrix (1-RDM) theory with the Hartree-Fock molecular orbitals and energies~\cite{sajjan2018current, raeber2019current}, we compute the intrinsic molecular conductance of both molecules in two cases: (1) conductance in the plane of the molecule and (2) conductance around the molecular ring as potentially driven by a magnetic field through the molecule's center.

We find for both cyclo[18]carbon and B\textsubscript{9}N\textsubscript{9} that conductance in the plane of the molecule is greater than conductance around the ring.  Furthermore, cyclo[18]carbon is slightly more conductive than B\textsubscript{9}N\textsubscript{9} for both (1) and (2). The intrinsic conductance provides information about the intrinsic ability of a molecule to support conductance that does not consider the resistance from the junction or lead.   Intrinsic conductance is useful for assessing whether a molecule is a potentially good molecular conductor or insulator prior to optimization of the lead and/or junction. The conductance per molecular orbital is also obtained, providing insight into how the orbitals---their energies and densities---drive the conduction.

\begin{figure}
\centering
\subfloat{\includegraphics[width=0.5\textwidth]{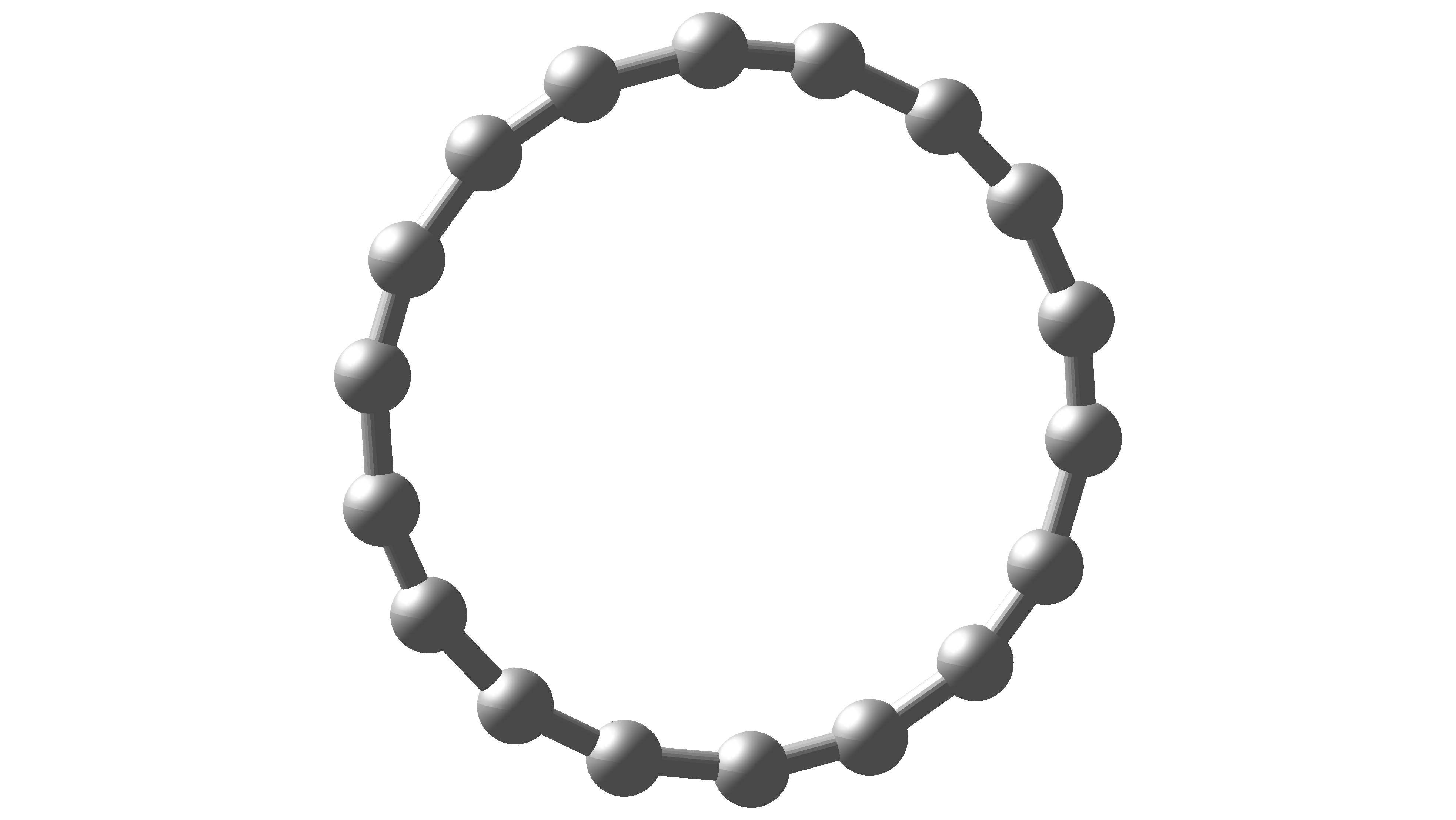}} \\
\subfloat{\includegraphics[width=0.5\textwidth]{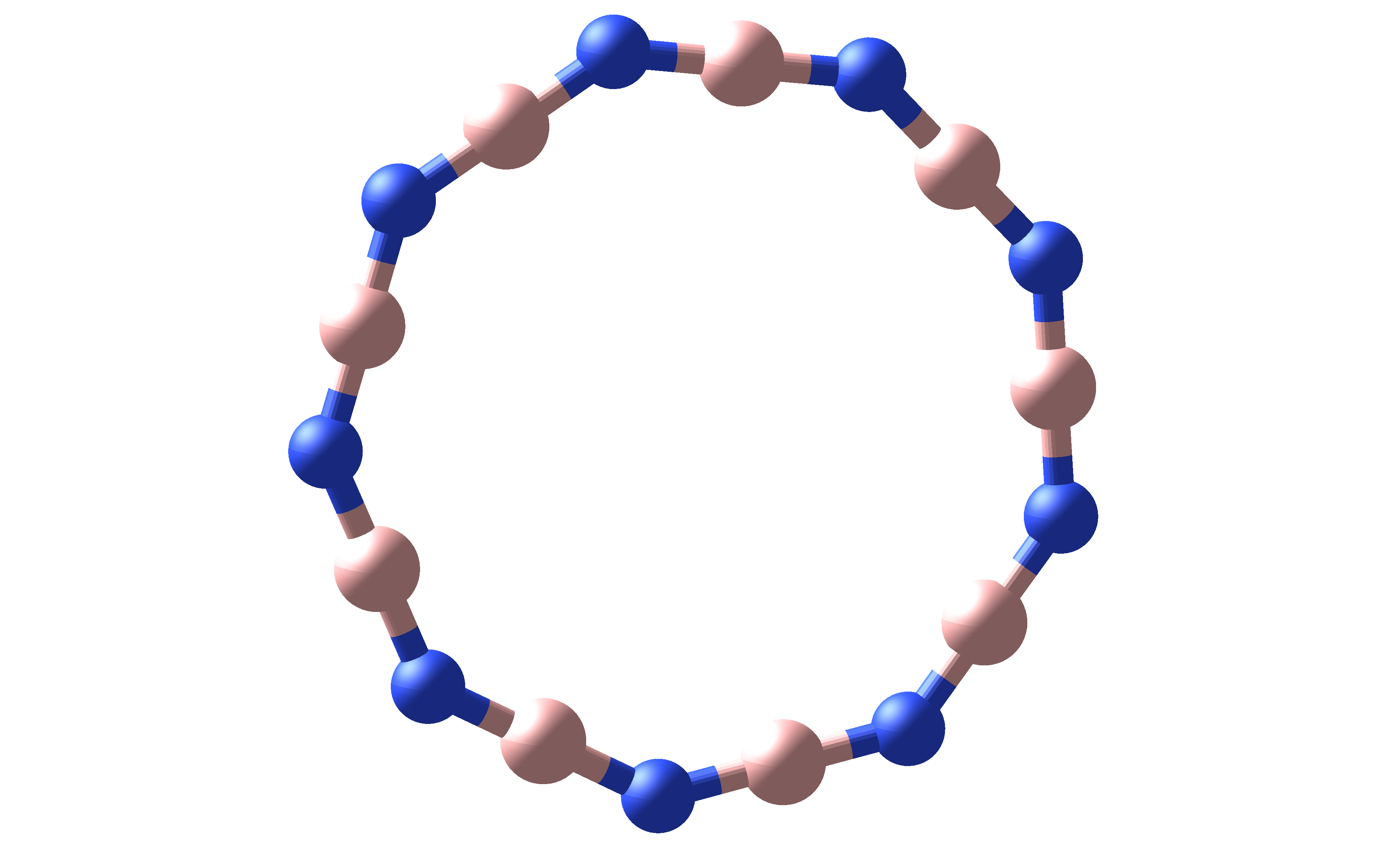}}
\caption{Ball and stick representations of C\textsubscript{18} (top) and B\textsubscript{9}N\textsubscript{9} (bottom)} \label{fig:geometries}
\end{figure}

\section{Theory}

The idea of creating single molecule circuit elements has been of scientific interest for nearly fifty years ~\cite{ar1974}, with a great deal of progress being made in the past twenty years~\cite{chen1999,tour2001,batra2013}. Traditional theories of molecular conductivity add a voltage to the molecule to compute the current~\cite{xiang2016, hsu2017, taylor2001, damle2001,diventra2001, xue2002, brandbyge2002,varga2011,rothman2010,hoy2017}.  In the current-constrained RDM theories the current is added as a constraint to compute the reorganization energy and the voltage~\cite{sajjan2018current, raeber2019current}. This change in paradigm allows us to compute an intrinsic conductance (or resistance) for each molecule that depends only upon a molecule's energetic response to the imposed current constraint, as well as its intrinsic properties such as length and polarization.  The intrinsic resistance reflects key qualitative features of molecular conductivity, features that are determined mainly by the electronic properties of the molecule.

To define a 1-RDM theory of molecular conductivity, we begin with the variational principle in the absence of electron transport. The energy of the system, given by
\begin{equation}
E = \text{Tr}(^{1}K ^{1}D),
\end{equation}
where $^{1}K$ is the matrix form of the one-electron reduced Hamiltonian, which we approximate  as the Fock matrix and $^{1}D$ is the 1-RDM. This energy is minimized subject to ensemble $N$-representability conditions on the 1-RDM known as the 1-positivity constraints, which ensure that the 1-RDM generated in the minimization arises from the integration of at least one $N$-electron ensemble density matrix~\cite{coleman1963, coleman2000, mazz2012sec, piris16, carlos16, schilling18}. These conditions have the form,
\begin{align}
^{1}D &\succeq 0 \\
^{1}Q &\succeq 0
\end{align}
where $^{1}D$ and $^{1}Q$ are the one-particle and one-hole reduced density matrices and the symbol $\succeq$ indicates that they must remain positive semidefinite, that is, they must have non-negative eigenvalues. These conditions are equivalent to the Pauli exclusion principle, which restricts the occupation of each spin orbital to lie between 0 and 1. This method of energy minimization is a special type of convex optimization known as semidefinite programming (SDP)~\cite{mazz2004, zhao2004, mazz2011, mazz2012sec}.

To set the current for the system, we define the one-electron current matrix as the matrix representation of the one-electron gradient in a specified direction ${\hat \kappa}$, given by
\begin{equation}
^1 J^p_q = \frac{1}{L} \int_{dr} \phi_p(r)(\nabla \cdot \hat{\kappa}) \phi_q (r) dr,
\end{equation}
where $L$ is the length of the molecule, $r$ represents the electronic coordinates, $\hat{\kappa}$ is the vector direction of the current, and $\phi_p$ are the molecular orbitals~\cite{sajjan2018current}. We add this to the energy minimization by requiring that
\begin{equation}
\label{eq:J}
\text{Tr}(^{1}J \ \text{Im}(^{1}D)) = I,
\end{equation}
where $I$ is the current.

In order to investigate the conductivity in the ring rather than across it, changes are made to the form of the one-electron current matrix. For the ring current calculations,
\begin{equation}
^1 \Tilde{J}^p_q = \bigg(\sum_i^M L^{-1} \int_{dr} \phi_p(r)(\nabla \cdot \hat{\kappa_i}) \phi_q (r) dr \bigg) M^{-1},
\end{equation}
where $M$ is the number of atoms in the ring and  $\kappa_i$ is the unit vector tangent to the center of each atom.


\section{Results and Discussion}

\subsection{Electronic Structure}

Molecular geometries of C\textsubscript{18} and B\textsubscript{9}N\textsubscript{9} as shown in Fig.~\ref{fig:geometries} are obtained by optimization at the v2RDM-CASSCF/cc-pVDZ level of theory~\cite{dunning1989gaussian, mazziotti2004first,mazz2004, gidofalvi2008active, mazz2011, mazz2012sec} with an active space of 18 electrons in 18 orbitals, implemented in the Quantum Chemistry Package~\cite{qcp} in Maple~\cite{maple}. The v2RDM method predicts the experimental polyynic structure, with alternating carbon-carbon bond lengths of 1.50~\AA \ and 1.22~\AA. The calculated structure of B\textsubscript{9}N\textsubscript{9} agrees well with the DFT structure from recent previous work~\cite{pichierri2020boron}, with a boron-nitrogen bond length of 1.32~\AA.  The Cartesian coordinates of the optimized geometries are provided in the Supporting Information.

\begin{figure}
\centering
\subfloat{\includegraphics[width=0.2\textwidth]{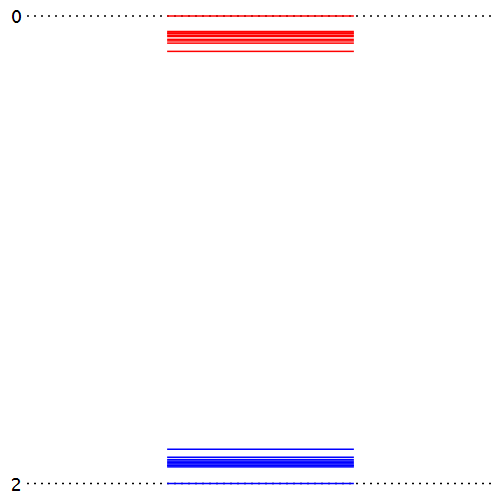}}
\subfloat{\includegraphics[width=0.2\textwidth]{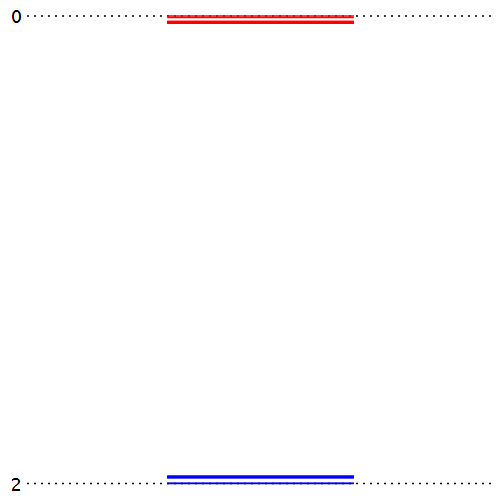}}
\caption{NO occupations of C\textsubscript{18} (left) and B\textsubscript{9}N\textsubscript{9} (right)} \label{fig:NO occupations}
\end{figure}

\begin{figure}
\includegraphics[width=0.5\textwidth]{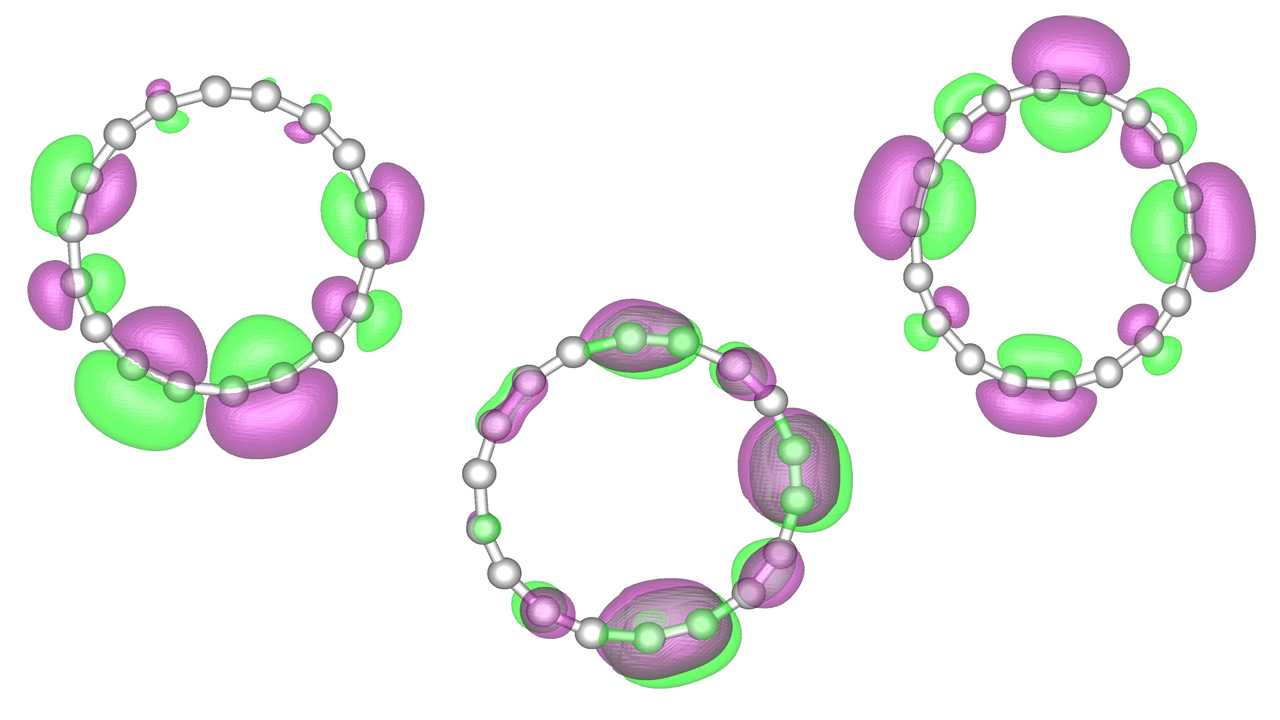}
\caption{A sample of the active space molecular orbitals for C\textsubscript{18}}\label{fig:C18 MOs}
\end{figure}

\begin{figure}
\includegraphics[width=0.5\textwidth]{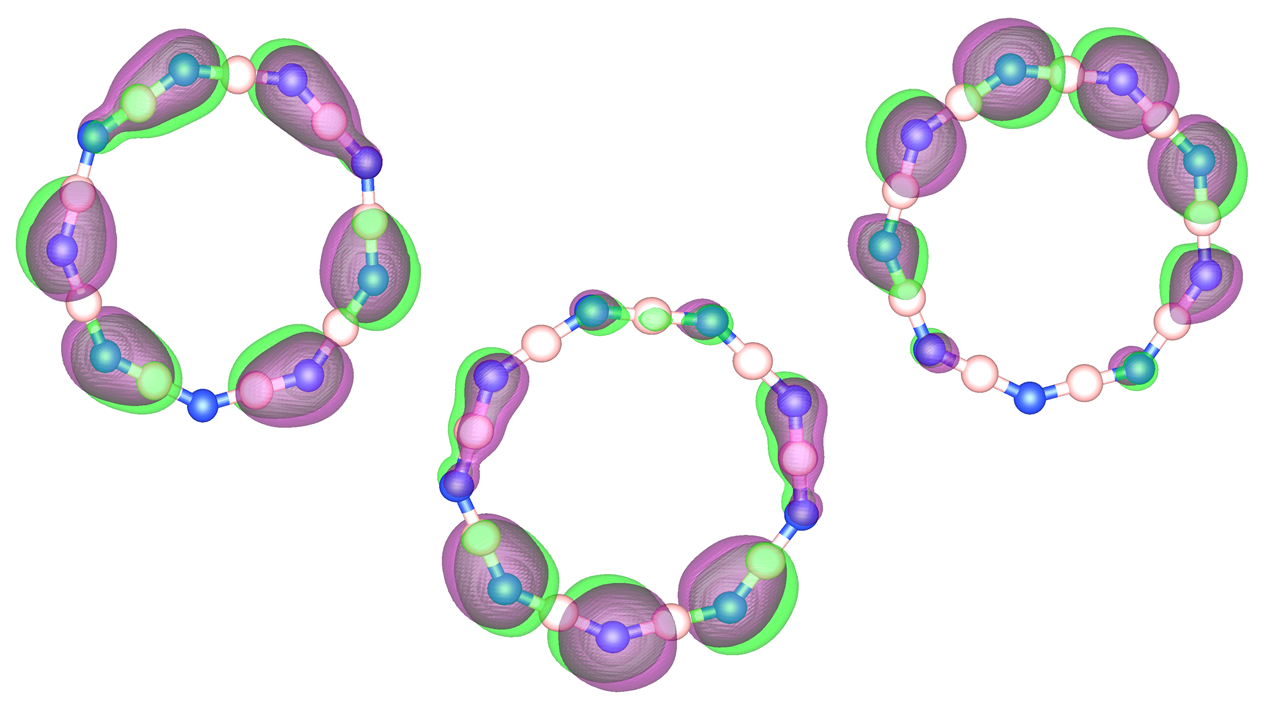}
\caption{A sample of the active space molecular orbitals for B\textsubscript{9}N\textsubscript{9}}\label{fig:B9N9 MOs}
\end{figure}

One of the key features of C\textsubscript{18} is its dual $\pi$ system, which is well represented in the v2RDM results, with $\pi$-type active molecular orbitals which lie either in the plane of the molecule or perpendicular to it. A representative subset of the active orbitals is shown in Fig.~\ref{fig:C18 MOs}. Our calculations indicated that this dual $\pi$ system is shared by B\textsubscript{9}N\textsubscript{9}, with a representative subset shown in Fig.~\ref{fig:B9N9 MOs}.

We also find that these molecules do not have a high level of static correlation. The natural orbital occupations in the active space, as shown in Fig.~\ref{fig:NO occupations}, do not diverge much from a completely filled (2) or empty (0) state. The occupation of the highest occupied natural orbital (HONO) is 1.85 and the occupation of the lowest unoccupied natural orbital (LUNO) is 0.15. B\textsubscript{9}N\textsubscript{9} displays even less static correlation than C\textsubscript{18}, with a occupation of 1.97 in the HONO and occupation of 0.03 in the LUNO. The minimal static correlation in these molecules supports the computation of their molecular conductivity within a 1-RDM theory.

\begin{figure}[h]
\includegraphics[width=0.5\textwidth]{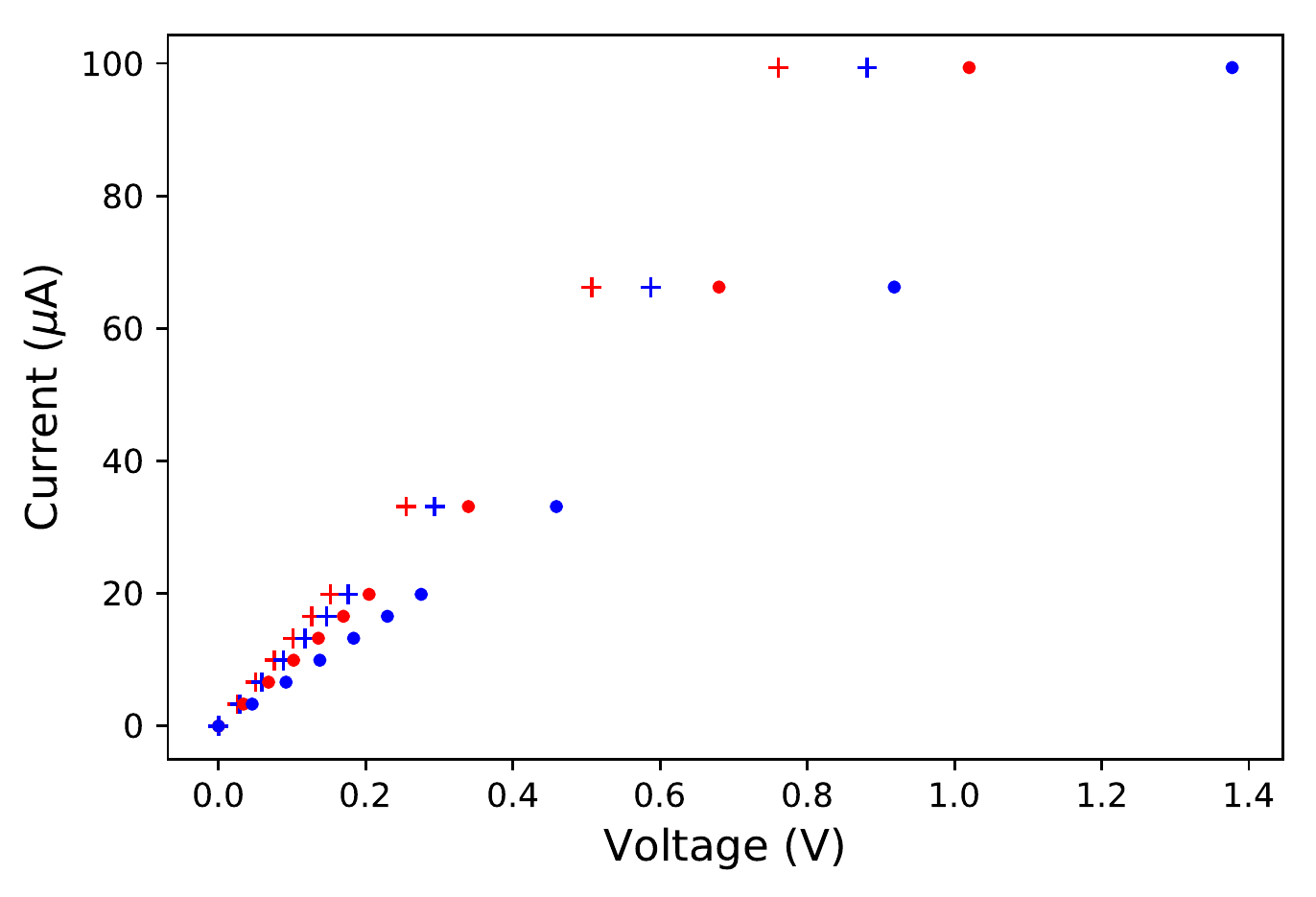}
\caption{Voltage plotted as a function of current for C\textsubscript{18} ring (red circle), C\textsubscript{18} in-plane (red plus), B\textsubscript{9}N\textsubscript{9} ring (blue circle), B\textsubscript{9}N\textsubscript{9} in-plane (blue plus).}\label{fig:iv curves}
\end{figure}


\subsection{Molecular Conductivity}

We investigate the conductance both around the molecular ring and in the plane of the molecule for cyclo[18]carbon and B\textsubscript{9}N\textsubscript{9}. All molecules are centered at the origin and the vector direction $\hat{\kappa}$ of the current is set as the x-axis for the calculation of the in-plane current and as described in the theory section for the ring current. We generate the Hamiltonian using the Fock matrix from a minimization of the Hartree-Fock energy of each molecule in the cc-pVDZ basis set.  Since we are interested in a qualitative description of the conductivity which arises only from the molecule itself, we compute the intrinsic conductance of the molecule, omitting the thiol linkers and metal leads present in an experimental molecular junction.

As shown in Fig.~\ref{fig:iv curves} we find that cyclo[18]carbon is the more conductive of the two molecules, with a ring conductance of 97.42 $\mu S$ and a in-plane conductance of 130.64 $\mu S$ compared to a ring conductance of 72.14 $\mu S$ and a in-plane conductance of 112.73 $\mu S$ for B\textsubscript{9}N\textsubscript{9}. The lower conductance of B\textsubscript{9}N\textsubscript{9} is expected based on its larger energy gap between the highest occupied molecular orbital (HOMO) and the lowest unoccupied molecular orbital (LUMO), with a gap of 13.28~eV compared to the 2.97~eV gap for C\textsubscript{18}. While experimental conductance values for these molecules are not available, the conductance of cyclo[18]carbon does agree well with that recently calculated using an NEGF-DFT method~\cite{zhang2020diverse}. For both molecules, the in-plane conductance through the center of the molecule is greater than the ring conductance.

\begin{figure}
\centering
\subfloat{\includegraphics[width=0.5\textwidth]{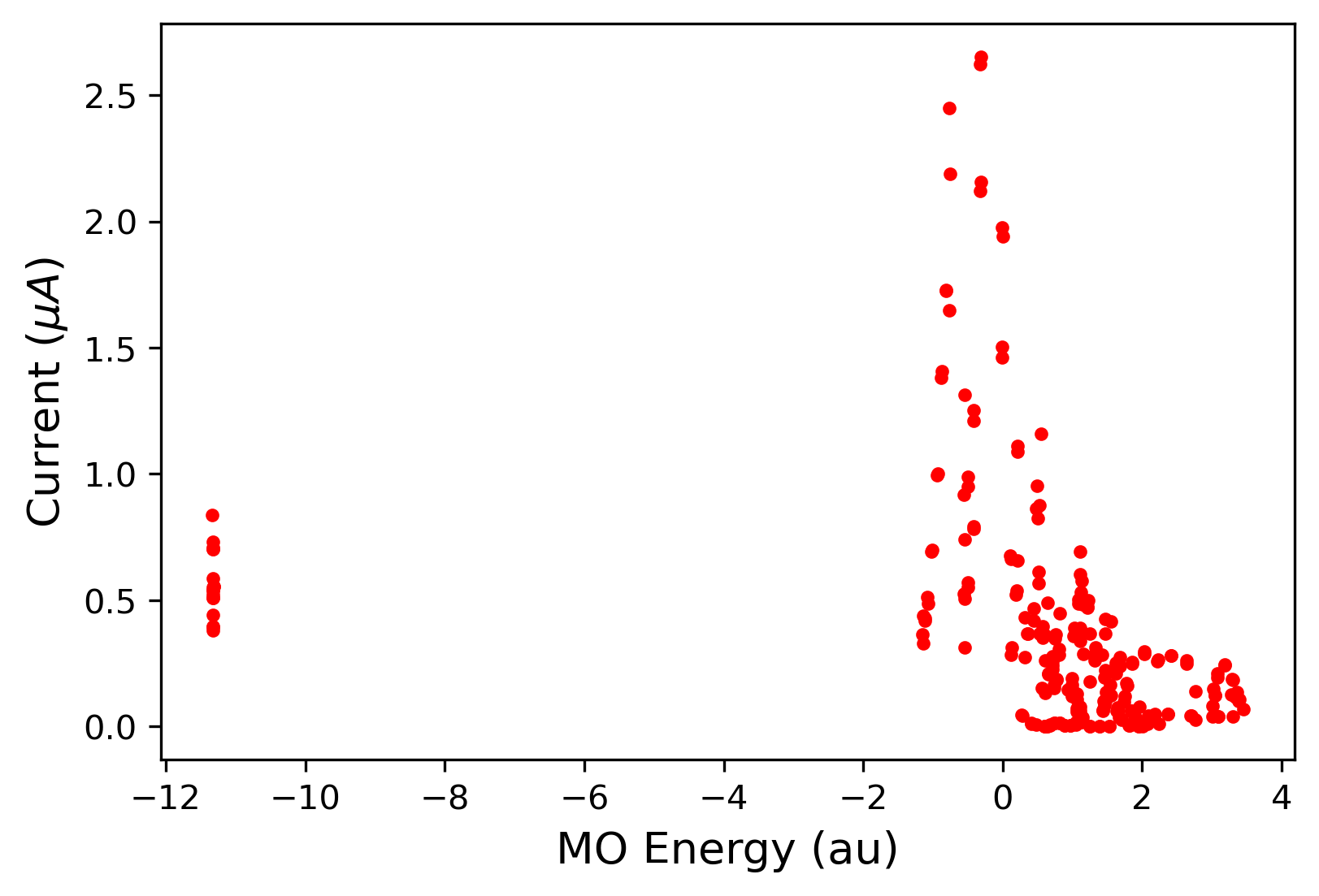}} \\
\subfloat{\includegraphics[width=0.5\textwidth]{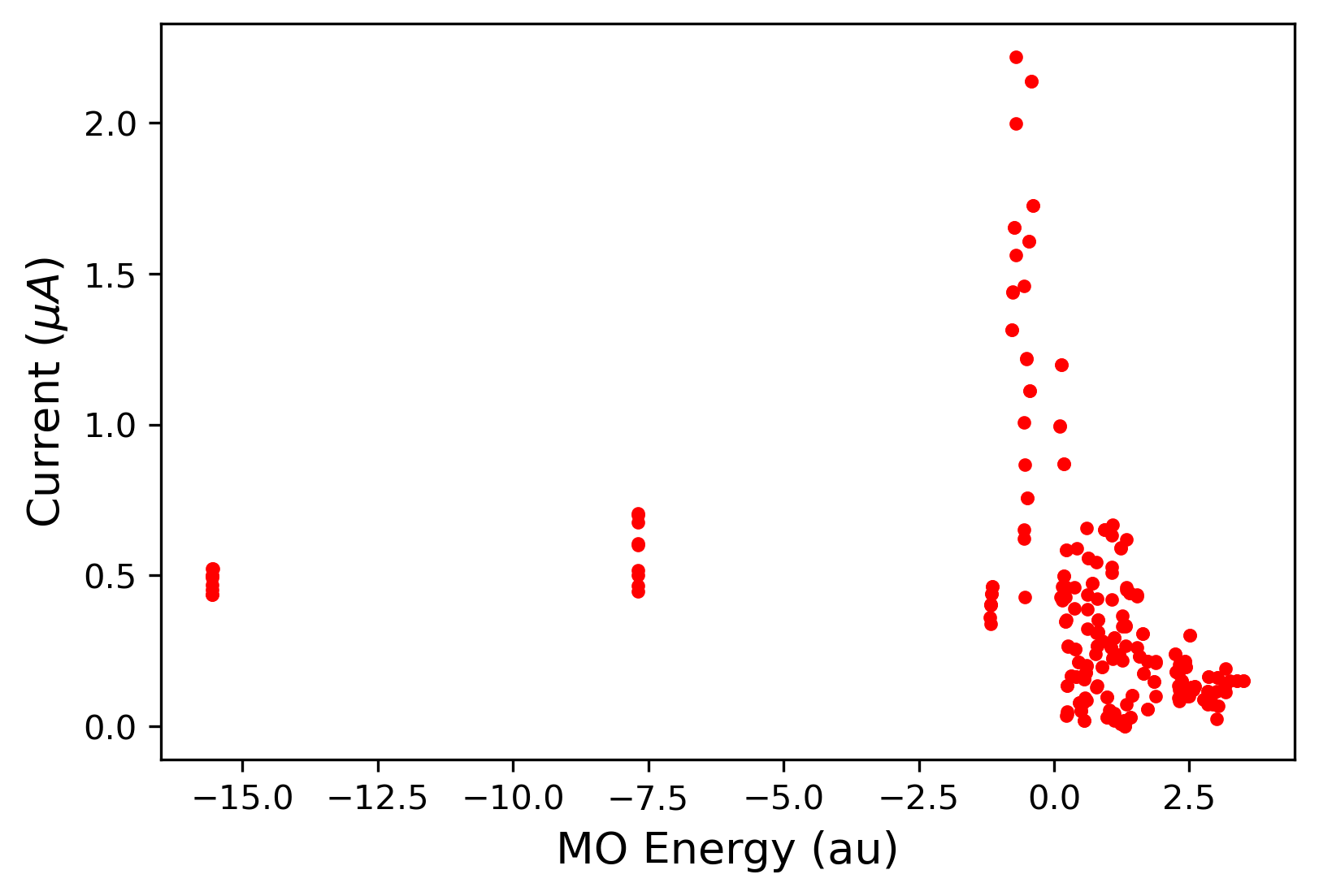}}
\caption{Current per MO as a function of MO energy for C\textsubscript{18} (top) and B\textsubscript{9}N\textsubscript{9} (bottom). In both cases a total of 100 $\mu A$ of current is imposed on the system. }
\label{fig:curr per mo}
\end{figure}

We are also able to investigate the conductance per molecular orbital for the molecules in question, as shown in Fig.~\ref{fig:curr per mo}. For both molecules a small number of molecular orbitals near the HOMO-LUMO gap are the most involved in conductance. Ten percent of the orbitals contain thirty-nine percent of the current in C\textsubscript{18} and thirty-seven percent of the current in B\textsubscript{9}N\textsubscript{9}. These orbitals are a mix of $\pi$ and $\sigma$ orbitals which are symmetric or nearly symmetric. The core orbitals and the high-energy virtual orbitals are much less involved in conductance than those near the Fermi level, with the core orbitals carrying on average an order of magnitude less current and the high-energy virtual orbitals carrying as much as four orders of magnitude less current. While implementations of many traditional theories like NEGF-DFT require the selection of an energy range for the conductivity, the current-constrained RDM theory predicts the energy (orbital) channels that support the molecular conductivity.

\section{Conclusions}

In the current-constrained 1-RDM theory the current is added as a constraint to compute the voltage, the opposite of traditional theories, which impose a voltage on the molecule to compute the resulting current.  As a result of this change in paradigm, we compute an intrinsic conductance for each molecule, a quantity which depends only upon a molecule's energetic response to the imposed current and its electronic structure. As shown in the results, the intrinsic conductance provides a good starting point for the determination of molecular species which may be useful for molecular electronic applications, as molecules without good intrinsic conductance will not be good experimental conductors.

The current-constrained RDM theory also allows for the study of facets of the conductivity which are not easily accessible using traditional theories, including the ring conductance and the orbital and energy localization of the current.  The direct computation of the 1-RDM allows us to construct the equivalent of a molecular-orbital diagram in molecular conductivity where the conductance of each orbital is presented as a function of energy.  Just as molecular-orbital diagrams provide an intuitive picture for molecular bonding, these molecular-orbital conduction diagrams provide an intuitive picture for molecular conductivity.

In the this work we investigate the electronic structure of cyclo[18]carbon and its boron nitride analogue using the v2RDM-CASSCF method.  We find that the molecular geometry  of C\textsubscript{18 }obtained from v2RDM optimization agrees with that found in a recent experiment, and that the key features of its electronic structure are well-described.  We apply the current-constrained theory for molecular conductivity to study the ring and in-plane conductivity of these molecules. We find that they have higher in-plane conductivity than ring conductivity and that C\textsubscript{18} is more conductive than B\textsubscript{9}N\textsubscript{9}.

\vspace{1cm}

\noindent {\bf Conflicts of Interest:} There are no conflicts of interest to declare.

\begin{acknowledgments}
AER acknowledges the support of the Department of Defense (DoD) through the National Defense Science and Engineering Graduate Fellowship (NDSEG) Program.  DAM gratefully acknowledges support from the ACS Petroleum Research Fund Grant No. PRF No. 61644-ND6 and the U. S. National Science Foundation Grant No. CHE-1565638.
\end{acknowledgments}
\bibliography{ring_conduct}
\end{document}